\documentclass[modern]{aastex701}

\usepackage{float}
\usepackage{xspace}

\graphicspath{{paper_figs_small/}} % 展開後にフォルダ名を変えるなら適宜調整

\newcommand{\bea}{\begin{eqnarray} }
\newcommand{\eea}{\end{eqnarray}}

\newcommand{\oiii}{[O{\scriptsize III}] }

\newcommand{\brg}{Br\,$\gamma$\xspace}

\newcommand{\cloudy}{C{\scriptsize LOUDY~}}

\begin{document}

%\title{Template \aastex v7.0.1 Article with Examples\footnote{Footnotes can be added to titles}}
\title{Broad-line Regions behind Haze: The Intrinsic Shape of the Br$\gamma$ Line and Its Origin in a Type 1 Seyfert Galaxy}

\author[0000-0002-8779-8486]{%
Keiichi Wada
}
\affiliation{Kagoshima University, Graduate School of Science and Engineering, Kagoshima 890-0065, Japan}
\affiliation{Ehime University, Research Center for Space and Cosmic Evolution, Matsuyama 790-8577, Japan}
\affiliation{Hokkaido University, Faculty of Science, Sapporo 060-0810, Japan}
%\correspondingauthor{Keiichi Wada}
\email[show]{wada@astrophysics.jp}

\author[0000-0002-7402-5441]{%
Tohru Nagao
}
\affiliation{Ehime University, Research Center for Space and Cosmic Evolution, Matsuyama 790-8577, Japan}
\affiliation{Kagoshima University, Graduate School of Science and Engineering, Kagoshima 890-0065, Japan}
\email{nagao@ehime-u.ac.jp}

\author[0000-0002-2125-4670]{%
Taro Shimizu
}
\affiliation{Max Planck Institute for Extraterrestrial Physics}
\email{simizu@mpe.mpa.de}

\author[0000-0002-5687-0609]{%
Daryl Joe D. Santos
}
\affiliation{Max Planck Institute for Extraterrestrial Physics}
\affiliation{De La Salle University, 2401 Taft Avenue, Malate, Manila, 0922 Philippines}
\email{dsantos@mpe.mpg.de}

\author[0000-0002-4569-9009]{%
Jinyi Shangguan
}
\affiliation{Kavli Institute for Astronomy and Astrophysics at Peking University}
\email{shangguan@pku.edu.cn}

\author[0000-0003-4949-7217]{%
Richard Davies}
\affiliation{Max Planck Institute for Extraterrestrial Physics}
\email{davies@mpe.mpa.de}

\begin{abstract}
The broad-line region (BLR) of active galactic nuclei is an essential component, yet its small size keeps its origin, structure, and kinematics uncertain. 
Infrared interferometry with VLTI/GRAVITY is now resolving the BLR-scale emission, with data for NGC~3783 consistent with a rotating, geometrically thick configuration. However, the processes shaping the spectra remain poorly constrained, and the cloud models are tuned phenomenologically rather than derived from first-principle predictions.
We address this by coupling three-dimensional radiation-hydrodynamic (RHD) simulations of gas around a supermassive black hole with radiative-transfer calculations using \cloudy, comparing the results to the SINFONI Br$\gamma$ profile of NGC~3783. We find that Br$\gamma$ arises from ionized gas in the surface of the rotating thin disk, with electron temperatures of approximately $T_e \approx 10^4$ K and number densities of 
$n_e \approx 10^8-10^{11}$ cm$^{-3}$. However, the intrinsic line profile produced by the RHD kinematics is narrower than observed and displays substructure.
An approximate treatment of the electron scattering suggests that scattering in the surrounding diffuse ionized gas significantly broadens and smooths the intrinsic Br$\gamma$ profile, making it consistent with the observed profile. This scattering medium has an electron temperature of  $10^4 - 10^5$ K,  and a number density of $n \lesssim 10^8$ cm$^{-3}$. Although a best-fit viewing angle of $\approx 15^\circ$ is suggested, the scattered line is notably less sensitive to inclination than the intrinsic line. 
The observed BLR profiles may be understood as the intrinsic emission viewed through an electron-scattering haze, such that some spectral detail is plausibly redistributed rather than seen directly.
%This perspective may offer a potential resolution to the long-standing “cloud versus flow” debate concerning BLR structures, though it will likely be resolved through a more complete treatment of scattering processes and radiative fluid dynamics.

\end{abstract}

%\keywords{\uat{Galaxies}{573} --- \uat{Cosmology}{343} --- \uat{High Energy astrophysics}{739} --- \uat{Interstellar medium}{847} --- \uat{Stellar astronomy}{1583} --- \uat{Solar physics}{1476}}


\section{Introduction} \label{sec:intro}

Active galactic nuclei (AGNs) are extraordinarily luminous sources powered by the accretion of matter onto a central supermassive black hole (SMBH) \citep{LyndenBell1969, peterson_1997, padovani2017}. These objects exhibit characteristic broad emission-line spectra arising from high-density, high-velocity gas in the so-called broad-line region (BLR), which lies in the immediate vicinity of the SMBH \citep{Sulentic2000,Peterson2006}. A detailed understanding of the BLR’s structure and kinematics is crucial both for precise SMBH mass determinations and for elucidating the mechanisms of accretion and outflow in AGNs.

However, the BLR is extremely compact—even the outer part of the BLR is a factor of three smaller than the dust sublimation radius \citep{Netzer2013, koshida2014}—and thus its physical origin, detailed structure, and the state of its constituent gas clouds (the line-emitting medium) have remained observationally unresolved for decades. Indeed, the very nature of the BLR is often posed as an open question, and numerous models have been proposed to account for its various aspects \citep[e.g.,][]{mineshige1990, Emmering1992_EBS92_MHDwind, chiang-murray1996,Wang2017_TidalDustClumps, Czerny2019}, {although one can infer partially 
the structures and geometry of BLRs using velocity-resolved reverberation mapping \citep[e.g.,][and references therein]{bentz2009, Li_2024}.}

Two broad classes of models have been proposed. In one, the BLR consists of a very large number of discrete, nearly noninteracting ionized clouds \citep{Krolik1981TwoPhase,Rees1989SmallDense,Baldwin1995LOC}. 
In the other, the emission arises from a smooth hydrodynamic flow—such as a rotating disk, a disk wind, and outflows \citep{Murray1997DiskWind,Proga2000LineDrivenWinds}. 
Within the cloud picture, it has been argued that sufficiently many high-density clouds can naturally produce the observed smooth, featureless line profiles \citep{Baldwin1995LOC}. 
High-S/N, high-resolution spectroscopy shows no obvious ``granularity" in the profiles, implying extremely large numbers of emitting elements—typically $N_{\rm cloud} \gtrsim 10^{6}$-$10^{8}$—which strains the simple discrete-cloud interpretation 
\citep{Arav1997Mrk335,Arav1998NGC4151,laor2006ClumpedSmooth}. However, the basic physics of how such ionized clouds form and are maintained in a quasi-steady state remains unclear \citep{Krolik1981TwoPhase,Rees1989SmallDense}. 
Moreover, if significant cloud-cloud interactions (e.g., collisions) occur, the ensemble may in any case relax toward a geometrically thin, rotation-supported configuration that becomes observationally difficult to distinguish from smooth-flow models.
In the smooth-flow picture, by contrast, it is comparatively straightforward to reproduce single-peaked, smooth profiles, and it can be explained by
physically reasonable phenomena—Keplerian rotation in a disk and various forms of radiation-driven winds \citep{Murray1997DiskWind,Proga2000LineDrivenWinds,Matthews2020StratifiedDiscWind}. That said, a purely optically thin, axisymmetric disk generically produces double-peaked lines, whereas such profiles are observed only in a minority of AGN \citep{Eracleous1994DoublePeaked,Strateva2003DoublePeaked,StorchiBergmann2003ApJ}. 

In recent years, infrared interferometry—most notably with the VLTI/GRAVITY instrument \citep{GRAVITY2017}—has enabled spatially resolved observations of the BLR, allowing its structure and kinematics to be probed directly \citep{GRAVITY2018, GRAVITY2020a, GRAVITYCollab_2024_AA_684_A167}. This breakthrough has opened a new avenue for addressing long-standing questions about the origin of BLRs. NGC~3783 is one of the brightest and most active nearby type-1 AGNs \citep{Bentz2021, GRAVITY2021a, GRAVITY2021b}, and its BLR has been studied in detail through near-infrared observations with VLTI/GRAVITY and SINFONI. For example, \citet{GRAVITY2021a} found that the \brg line constrains the geometry and dynamics of the BLR, revealing a rotating, geometrically thick disk on sub-parsec scales. High-resolution SINFONI spectra of \brg provide detailed line profiles, and rotating, thick-disk models reproduce the observed profiles remarkably well. Nevertheless, significant uncertainties persist in our understanding of the physical processes that shape the BLR’s emission-line spectra, because the BLR “cloud” model—its size, thickness, internal structures, and kinematics—is not necessarily based on first-principles physics.

A comprehensive understanding of these complex phenomena requires coupling radiation-hydrodynamic (RHD) simulations with detailed radiative-transfer calculations. Such radiative-transfer calculations, based on physics-driven hydrodynamic models, can capture the intricate physical conditions. The resulting model spectra can be directly compared with observations. \citet{Wada2012} and subsequent related papers found that radiative feedback forms a fountain-like flow of gas (“radiation-driven fountain”) around an AGN on scales of tens of parsecs, naturally explaining the type-1/type-2 dichotomy of Seyfert galaxies \citep{schartmann2014, wada2015obscuring}. A model snapshot from \citet{Wada2016} was tested with post-processed radiative-transfer calculations, reproducing the observed spectral energy distribution (SED) of the Circinus galaxy when viewed nearly edge-on. \citet{P1} and \citet{P2} performed non-local thermodynamic equilibrium (non-LTE) line-transfer calculations of submillimeter molecular (CO) and atomic ([CI]) lines. 
\citet{P4} and \citet{Baba2024} calculated CO emission and absorption lines against the dust emission from the AGN. 
 In addition, \citet{M22} performed radiative transfer of near-infrared CO rovibrational lines and investigated the detectability of molecular inflows and outflows in the torus. The multiphase structures of the radiation-driven fountain on parsec scales in the central region of the Circinus galaxy were resolved by ALMA \citep{izumi2023}. \citet{P5} computed the X-ray polarization and compared it with a recent observation of the Circinus galaxy with the Imaging X-ray Polarimetry Explorer (IXPE). \citet{buchner2021} and \citet{Ogawa+22} calculated the X-ray spectrum based on radiation-driven fountain models and succeeded in reproducing the observed hard X-ray spectrum as well as the iron lines of Seyfert galaxies.

For optical spectra of photo-ionized gas on 10-pc scales, \citet{wada2018NLR} focused on the bipolar outflow of the radiation-driven fountain and performed pseudo-three-dimensional radiative-transfer calculations using \cloudy \citep{ferland2017}. They found that the observed biconical narrow-line region, typically traced by \oiii $\lambda 5007$, can originate in radiation-driven outflows, and that the line ratios (i.e., the Baldwin-Phillips-Terlevich diagram \citep{Baldwin1981_BPT_PASP}) in the outflows are consistent with those observed in Seyfert galaxies. On BLR scales, \citet{Wada2023} investigated the spectral properties of ionized gas exposed to radiation from an AGN with a $10^7 M_\odot$ SMBH and found significant time-dependent variations in the Balmer emission lines.

In this paper, we apply this approach to the interstellar medium (ISM) further inside the dust sublimation radius of Seyfert galaxies. Through this analysis, we aim to reconcile the kinematic and geometric features inferred from observations with the underlying physical processes that give rise to the \brg emission, thereby providing deeper insight into the intrinsic structure and physical state of the BLR. We take a snapshot of three-dimensional RHD simulations of the ISM around an SMBH (Section~2.1). The model focuses on the gas in the central 0.01 pc of NGC~3783 and attempts to reproduce the \brg line profile obtained by SINFONI \citep{GRAVITY2021a}. The emission lines and continuum from gas selected from the whole computational domain, based on density and temperature criteria (Section~2.2), are calculated by \cloudy. The spectra are integrated to obtain the observed spectra, accounting for the line-of-sight velocity of the gas for a given viewing angle (Section~3.1). We also investigate the effect of electron scattering when integrating the original spectra, using a simple prescription (Section~3.2). The numerical results and comparisons with observations are given in Section~3. In Section~4, we discuss the possible physical state of the scattering medium.  We also give brief comments on the polarization of the BLR emission. 
Conclusions and future work are given in Section~5.


\section{Radiative Hydrodynamic Simulations}\label{Section0201}
\subsection{Numerical Methods}
The numerical method is based on \cite{Wada2012}, which is a three-dimensional Eulerian hydrodynamic code with 
the radiative feedback from the central source. 
We solve the three-dimensional evolution of a gas disk around an SMBH.
 The simulations utilize a grid of $256^3$ points. A uniform Cartesian grid $(x,y,z)$ spans a region of $0.02 \ \mathrm{pc} \times \ 0.02 \ \mathrm{pc} \times \ 0.01\ \mathrm{pc}$ around an SMBH. 
This box size is large enough to cover the observationally suggested BLR region in NGC~3783 \citep{Bentz2021}.
The spatial resolution for the $x$ and $y$ directions is $7.8 \times 10^{-5} \ \mathrm{pc}$ and that for the $z$ direction is $3.9 \times 10^{-5} \ \mathrm{pc}$. 
We numerically solve the following equations:

\begin{eqnarray}
\frac{\partial \rho}{\partial t} + \nabla \cdot (\rho \bm{v}) & =& 0, \label{Equation01}\\
\frac{\partial (\rho \bm{v})}{{\partial t}} + \rho (\bm{v}\cdot \nabla)\bm{v}+{\nabla p} & = &
-\rho \nabla \Phi_{\mathrm{SMBH}} + \rho \bm{f}_{\mathrm{rad}}^{r}, \label{Equation02}\\
\frac{\partial (\rho E)}{\partial t} + \nabla \cdot [(\rho E+p)\bm{v}] & = &
 \rho \bm{v} \cdot \bm{f}_{\mathrm{rad}}^{r}
+ \rho \Gamma_{\mathrm{UV}}(G_0)  \label{Equation03} \\  \nonumber 
&+&  \rho \Gamma_{\mathrm{X}}  - \rho^{2} \Lambda(T_{\mathrm{gas}}) - \rho \bm{v} \cdot \nabla \Phi_{\mathrm{SMBH}}. 
\end{eqnarray}

In these equations, $\rho$ represents the gas density, $\bm{v}$ is the gas velocity, $p$ is the gas pressure, $\Phi_{\mathrm{SMBH}}$ is the potential of the SMBH, $\bm{f}_{\mathrm{rad}}^{r}$ is the radiative force, $E$ is the specific total energy, $\Gamma_{\mathrm{UV}}$ is the heating due to the photoelectric effect, $\Gamma_{\mathrm{X}}$ is the X-ray heating rate, and $\Lambda$ is the cooling function. 
The self-gravity of the gas is ignored. 
In Equation \hyperref[Equation02]{Equation~2}, we adopted the following time-independent potentials of the SMBH: $\Phi_{\mathrm{SMBH}}(r) = -{GM_{\mathrm{SMBH}}}/{(r^2 + b^2)^{\frac{1}{2}}}, 
$
where $M_{\mathrm{BH}} = 2.83 \times 10^7 M_{\odot}$, which is suggested for NGC~3783 \citep{GRAVITY2021a}. To avoid numerical errors and small time steps near the innermost numerical grid cells, the gravitational potential is smoothed for $b = 5 \times 10^{-5} \ \mathrm{pc}$. 

We assumed an Eddington ratio of 0.1 \citep{GRAVITY2021a}. We set a sink boundary in the central eight grid cells to mimic the accretion to the SMBH. However, we keep the luminosity of the central source constant.
We examined the radial component of the radiation pressure:
$
\bm{f}_{\mathrm{rad}}^r = \int {\chi_{\mathrm{T}}\bm{F}_{\nu}}/{c} d\nu,
$
with the radial flux
$
\bm{F}_{\nu}(r) = {L_{\nu}(\theta) e^{-\tau_{\mathrm{T}}}}/({4\pi r^{2}} ) .
$
Here, $\chi_{\mathrm{T}}$ represents the total mass extinction coefficient due to Thomson scattering and $\bm{F}_{\nu}(r)$ is the radial component of the flux at a given radius with $\tau_{\rm T} =\int \chi_{\rm T} \rho ds$. The optical depth $\tau_{\rm T}$ is computed at each time step along a ray originating from the central source at every grid point; that is, $256^{3}$ rays are traced within the computational domain. 
{ We do not solve the internal structure of the source, assuming a point 
source with a non-isotropic radiation field.
The ultraviolet flux is assumed to be $L_\nu(\theta) \propto \cos \theta (1+2\cos \theta)$ \citep{netzer1987},
where $\theta$ denotes the angle from the rotational axis ($z$-axis).
The X-ray radiation is assumed to be originated in the halo of the accretion disk and spherically symmetric  \citep{netzer1987, xu2015}.
We obtain the column density between the source and each grid point by
the ray-tracing method.
Therefore, the radiation can be strongly attenuated for 
high-density regions, particularly in the direction of the disk.}

In Equation \ref{Equation03}, we used a cooling function, $ \Lambda(T_{\mathrm{gas}}) $, for the X-ray and UV irradiated nuclear gas \citep{Meijerink2005}. The photoelectric heating term $\Gamma_{\rm UV}$ does not affect the thermal structure of the ionized gas, which is the target in this paper.
 We set the temperature floor for $n \ge 10^{12}$ cm$^{-3}$, since we cannot follow the realistic 
thermal state for such very dense gas without solving the self-gravity of the gas and the radiative transfer in the optically thick gas.
However, such very dense gas is present only in a thin layer at the midplane ($z = 0$) of the disk, and it does not affect the BLR spectra.

We also account for X-ray heating, including interactions between high-energy secondary electrons and thermal electrons (Coulomb heating), as well as H$_{2}$ ionization heating in both the cold and the warm gas phases \citep{Maloney1996, Meijerink2005}. The Coulomb heating rate $\Gamma_{\mathrm{X,c}}$ for gas with a number density $n$ is given by
$
\Gamma_{\mathrm{X,c}} = \eta n H_{\mathrm{X}},
$
where $\eta$ represents the heating efficiency \citep{Meijerink2005}, $H_{\mathrm{X}} = 3.8 \times 10^{-25} \xi \ \mathrm{ergs} \ \mathrm{s}^{-1}$ is the X-ray energy deposition rate, and the ionization parameter $\xi = 4\pi F_{\mathrm{X}}/n = L_{\mathrm{X}} e^{-\int \tau_{\nu} d\nu}/n r^{2}$ with 
the X-ray luminosity $L_{\mathrm{X}}$.
For an optically thin hot gas with $T \gtrsim 10^{4} \ \mathrm{K}$, the effects of Compton heating and X-ray photoionization heating are included. The net heating rate \citep{Blondin1994} is approximately 
$
\Gamma_{\mathrm{X,h}} = 8.9\times 10^{-36} \xi (T_{\mathrm{X}} - 4T) +1.5\times 10^{-21} \xi^{1/4} T^{-1/2}(1-T/T_{\mathrm{X}}) \ \mathrm{erg} \ \mathrm{s}^{-1} \ \mathrm{cm}^{3},
$
with the characteristic temperature of the bremsstrahlung radiation $T_{\mathrm{X}} = 10^{8} \ \mathrm{K}$.

\begin{figure}[H]
 \centering
 \includegraphics[width=0.6\linewidth]{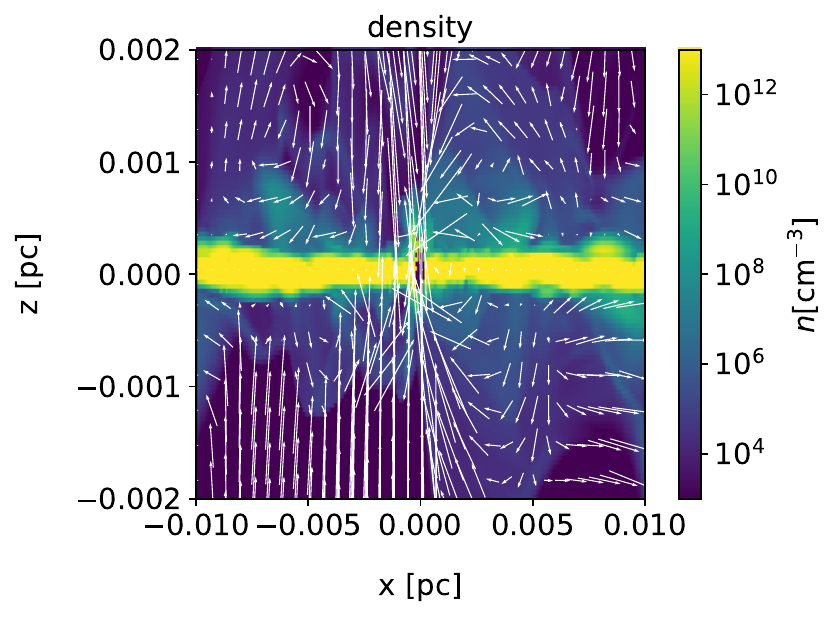} % original: aag0134_panel2.pdf
 \includegraphics[width=0.6\linewidth]{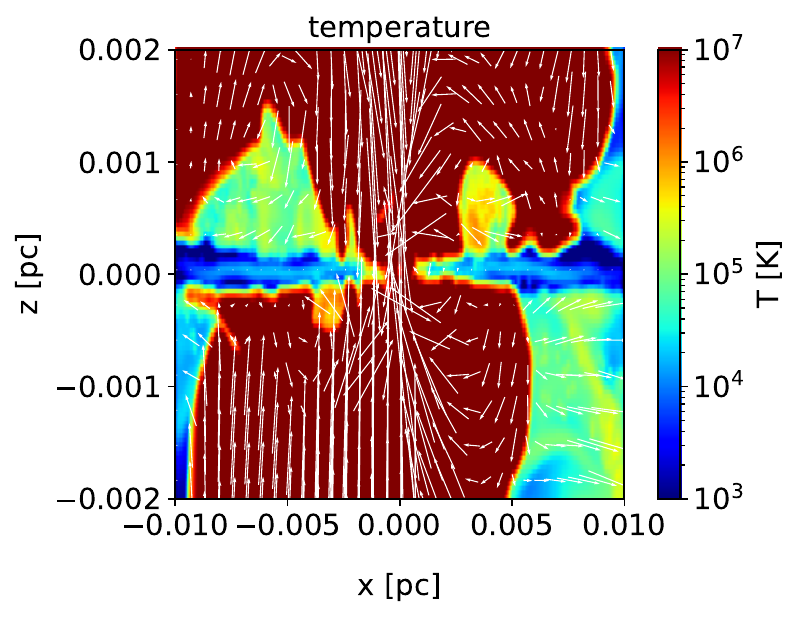} % original: aag0134_panel3.pdf
 \caption{Density slice and temperature slices with velocity vectors in the $x-z$ plane at $y 	= 0$ pc. }
 \label{fig:hydro}
\end{figure}

\subsection{Hydrodynamic Results and Candidate Cells of the BLR}
We evolve an axisymmetric, rotationally supported thin disk with a uniform density profile ($n = 10^{15}\, {\rm cm}^{-3}$, a thickness of $6.25 \times 10^{-4} \ \mathrm{pc}$) until the system becomes quasi-stable.
Figure \ref{fig:hydro} shows density and temperature slices of a snapshot that is used for the spectrum calculation (Section3). Note that
the plots are enlarged in the vertical direction.
The disk is almost axisymmetric, and geometrically thin with diffuse halo gas. Since the Eddington ratio of the central source is 
0.1 and there is no dust in this region, we do not see bipolar outflowing gas as typically seen in the 
radiation-driven fountain for larger scales \citep{Wada2016, wada2018NLR}; it rather forms an infalling and circulating gas with 
the density $ n \lesssim 10^8 {\rm cm}^{-3}$. The temperature of the dense disk is $10^4 $ K. 
The temperature of the diffuse halo does not show a thermally steady structure, reflecting complicated velocity structures with shocks and 
radiative heating processes.

\begin{figure}[H]
 \centering
 \hspace*{-1cm}
 \includegraphics[width= 1.2\linewidth]{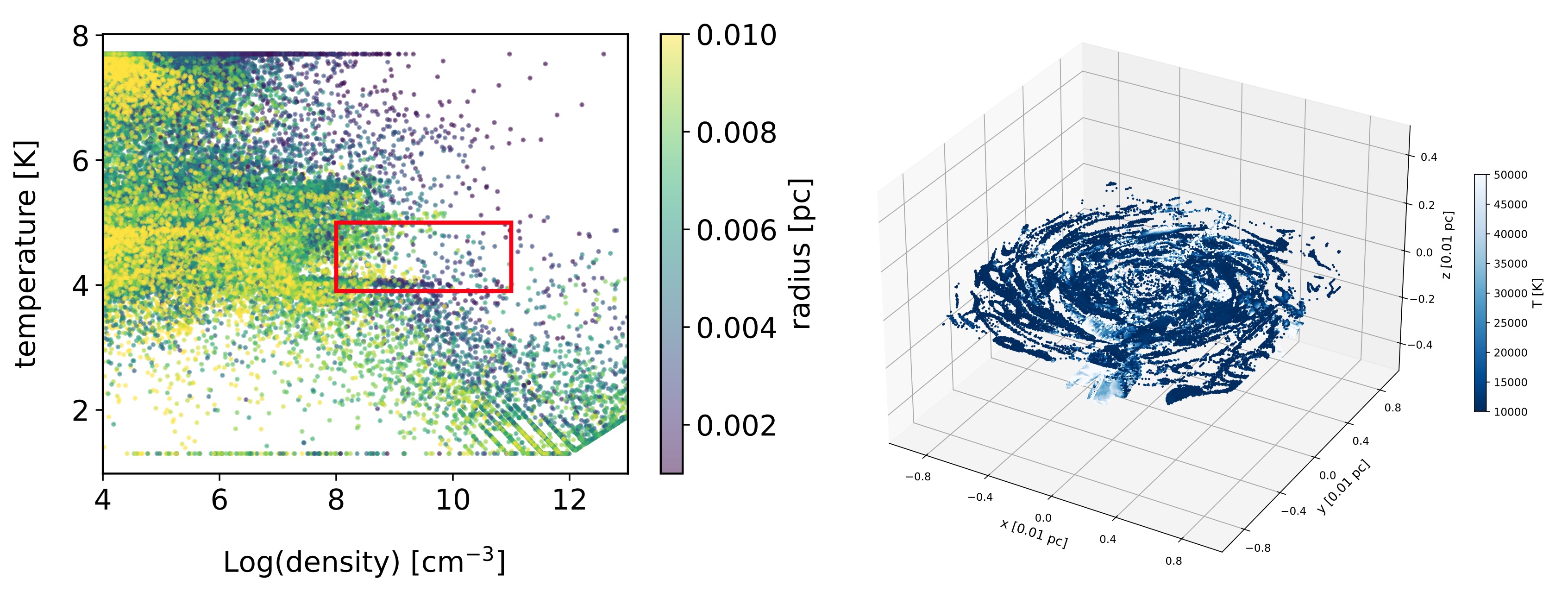}
 \caption{(Left) Density--temperature plot (colored by distance from the center). (Right) Three-dimensional distribution of the ``BLR candidate'' grid cells, which correspond to the cells in the red box of the density--temperature plot. }
 \label{fig:phase_dnT}
  % read_binary_phasediagram31.py (panel 1)
  % plot_BLR_data.py
\end{figure}

%\subsection{Criteria of BLR candidate cells}
In order to obtain the BLR spectrum, first, we selected ``potential candidates of BLR" from grid cells that satisfy 
the following criteria: for number density $10^8 \; {\rm cm}^{-3} \le n \le 10^{11} \; {\rm cm}^{-3} $, 
and for temperature 8000 K $ \le T \le 10^5$ K. These numbers are `typical' conditions of the BLR gas \citep{Peterson2006} and also suggested 
in local Seyfert 1 galaxies \citep{Schnorr-Muller2016}. 
In total, about $6\times 10^4$ grid cells are selected. 
In Figure \ref{fig:phase_dnT}, we show the selection conditions by a red box in the phase diagram and the spatial distribution of the
candidate cells.

\section{Radiative-transfer Calculations}

We calculate emission lines and continuum using \cloudy for the ``BLR candidate cells" selected by the criteria explained in Section2.2. 
Since it is not computationally feasible to run \cloudy simulations for all 60,000 grid cells, 
we randomly selected 2000 grid cells, and using their gas density, temperature, velocities, and positions, 
we run \cloudy to obtain spectra and integrate the output spectra, taking into account the Doppler shift for a given viewing angle. 
We repeated the same procedure for four different sets of 2000 grid cells to see the fluctuation due to the selection.
Ideally, one should solve the full three-dimensional radiative transfer including scattering; as a first step, however, we model these processes using a simplified, approximate approach (see Section3.2).

\subsection{Spectrum Calculations Using CLOUDY}

\begin{figure}[H]
 \centering
 \includegraphics[width=0.6\linewidth]{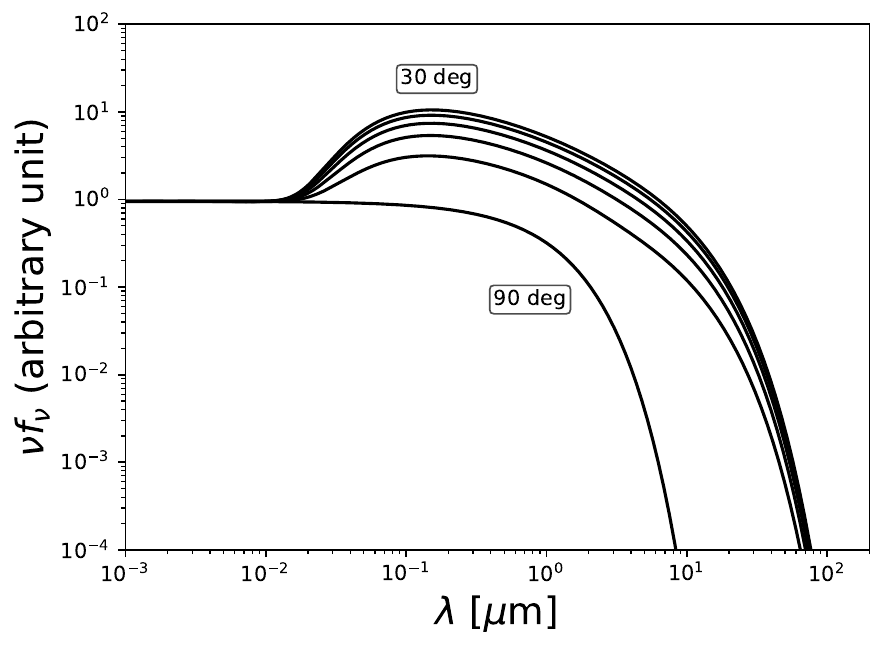} % original: input_SED.pdf
 \caption{The non-spherical SED of the central source for $i = 30,42, 54, 66, 78$, and 90$^\circ$.}
 \label{fig:sed_grid}
\end{figure}

We performed the spectral synthesis code \cloudy (version 23.01) \citep{cloudy23}
for the selected grid cells. 
The SED of the central source was derived from 
\cloudy's \textsf{AGN} command and is represented as
\begin{eqnarray}
\label{eqn:agncon}
F_\nu = \nu ^{\alpha _\mathrm{UV} } \exp \left( { - h\nu /kT_\mathrm{BB} } \right)\exp \left( { - kT_\mathrm{IR} /h\nu } \right)\cos{i} \nonumber\\
 + a\nu ^{\alpha_\mathrm{X} } \exp \left( { - h\nu /E_1 } \right) \exp \left( { - E_2 /h\nu } \right),
\end{eqnarray}
where 
$\alpha _\mathrm{UV} = -0.5$, $T_\mathrm{BB} = 10^5$~K, 
$\alpha_\mathrm{X} = -1.0$, $a$ is a constant that yields the X-ray-to-UV ratio $\alpha_\mathrm{OX} = -1.4$, 
 $kT_\mathrm{IR} = 0.01$~Ryd,
$E_\mathrm{1} = 300$~keV, $E_\mathrm{2} = 0.1$~Ryd, and $i$ is the angle from the rotational axis ($z$-axis).
The UV radiation (first term of Equation (\ref{eqn:agncon})), which derives from the geometrically
thin optically thick disk.
By contrast, the X-ray component (second term) was assumed to be isotropic  (Figure  \ref{fig:sed_grid}).
We assume that grains are sublimated (i.e., no grains) in the data used in C{\scriptsize LOUDY} ,
and the Solar metallicity is assumed. 
{The spectral resolution corresponds to $\sim$ 48 km s$^{-1}$ for $\lambda = 2.3\mu$m. }

Upon completion of all \cloudy  calculations, we “observed” the system along the line of sight by adopting a viewing angle $\theta_v$ (the angle between the line of sight and the disk’s rotational axis; $\theta_v=0$ corresponds to face-on). Because there is freedom in azimuth, we averaged the spectra over 16 azimuthal angles. Accordingly, for a given viewing angle, the spectra shown below are the result of integrating 2000 spectra for each of 16 azimuthal angles and then averaging over azimuth ($16\times2000$ spectra in total).

Each BLR “candidate cell” is assumed to be exposed to the AGN radiation. 
{ Here we assume optically thin 
conditions (though we do account for the angular dependence described in Figure \ref{fig:sed_grid}). Since the 
disk surface primarily forms the ionized BLR gas, this should be a 
reasonable zeroth-order approximation.}
In addition, to mimic radiative-transfer effects between the BLR cells, the output emission computed by \cloudy for a given cell is used as the input SED for the next \cloudy calculation; the \cloudy calculations are thus performed recursively.
{ However, the line transfer effects in the BLR and ``haze'' region are phenomenological in this sense, and 
fully incorporating electron scattering effects in 3D based on the 
hydrodynamic calculations is a task for future work (see also the next section).}

\subsection{Electron Scattering by the Ambient Gas and Radiative Transfer between "BLR Clouds"}
{
In this section, we explore how the \brg line profile changes through scattering by haze, i.e., an electron mist of ionized gas, in the BLR region.
\citet{kaneko1968} first pointed out that electron scattering can produce broadened, exponential wings in  emission lines of Seyfert-galaxies.
This idea was followed by \citet{Weymann1970} and \citet{Shields1981} in detail.}
The subsequent studies also showed that multiple Thomson scattering in a high-temperature intercloud medium ($T\sim10^4$-$10^7$ K) can affect line shapes and widths, and even produce small wavelength shifts \citep[e.g.,][]{Ferrara1993, Smith2004,laor2006, Peterson2006, Kim2007,Afanasiev2019}. \citet{laor2006} provided observational evidence for exponential H$\alpha$ wings in NGC 4395, attributing them to Thomson-scattering electrons at $T_e\sim1.1\times10^4$ K. More recently, JWST observations of “Little Red Dots” (LRDs)—compact, high-redshift AGNs—have shown that their broad-line profiles, especially at high redshift, are best explained by scattering through an optically thick, high-column-density electron scattering medium rather than by Doppler broadening alone \citep{Rusakov2025} \footnote{There is still a major ongoing debate concerning the origin of broad line profiles in high-redshift objects \citep[e.g.,][]{Juodzbalis2025JADES}.  }. In these objects, the inferred electron optical depths ($\tau_e=0.5$-$2.8$) and compact sizes (of order light-days) point to young SMBHs accreting in dense environments, reinforcing the view that electron scattering can dominate the formation of BLR line widths under high-density conditions.

\begin{figure}[H]
 \centering
 \includegraphics[width=0.6\linewidth]{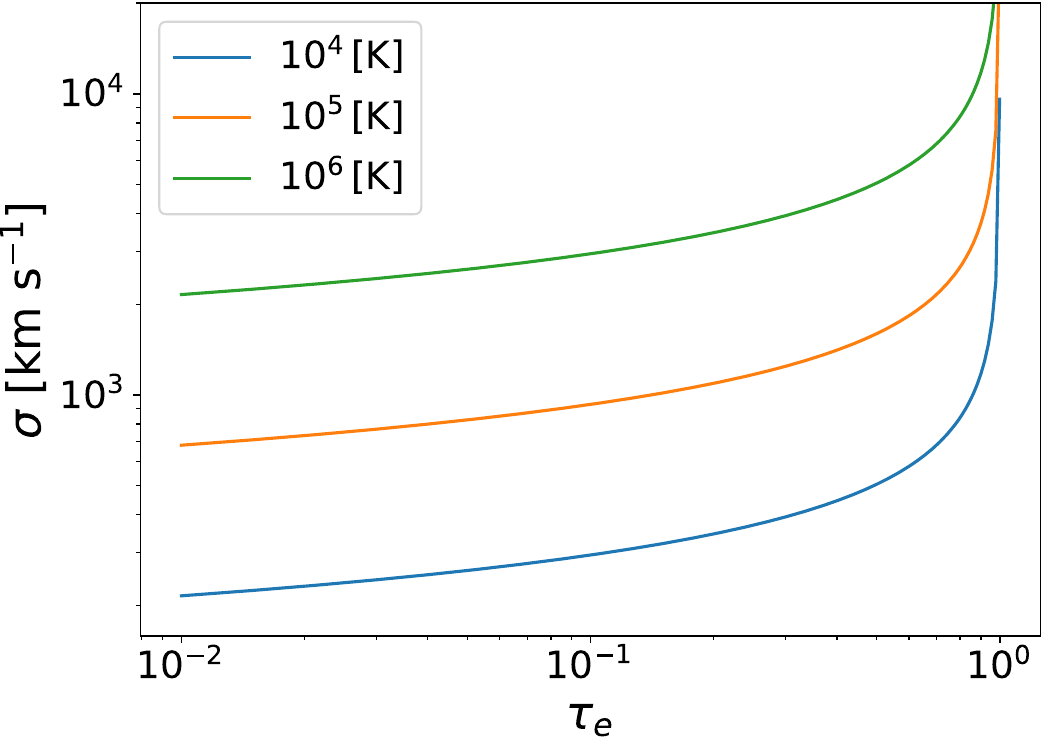} % original: sigma_all_tau_at_T100000.pdf
 \caption{The dispersion of the exponential tail by the electron scattering as a function of the optical depth $\tau_e$ (Equations \ref{eq: exp_tail} and \ref{eq: sigma}) for three electron temperatures $T_e$ \citep{laor2006}. }
 \label{fig:sigma-tau}
\end{figure}

Here, we take a simple prescription to evaluate the effect of the electron scattering on the line profile obtained by the manner explained
in Section~3.1 as follows. The original line spectrum $f_0(\lambda)$ for each ``BLR candidate'' cell calculated by \cloudy is assumed to have an exponential tail due to the scattering, following the fitting formula in \citet{laor2006}:
 \begin{eqnarray}
 f(v) = f_0 (\lambda) e^{-v/ \sigma},
 \label{eq: exp_tail}
 \end{eqnarray}
with the dispersion $\sigma$: 
 \begin{eqnarray}
 \sigma = 428 \, T_4^{1/2} (\ln \tau_e^{-1})^{-0.45} \; {\rm km}\, {\rm s}^{-1}, 
 \label{eq: sigma}
 \end{eqnarray}
 with the optical depth for the electron scattering $\tau_e$ and the electron temperature of the hot gas $T_4 \equiv T_e/10^4\, {\rm K}$. 
 Figure \ref{fig:sigma-tau} shows $\sigma$ as a function of $\tau_e$ for three electron temperatures ($T_4 = 1, 10$, and 100).
Then, all the line spectra for 2000 BLR candidate cells obtained by Equation~(\ref{eq: exp_tail}) are integrated, taking into account the Doppler shift for a given viewing angle.

\begin{figure}[H]
 \centering
 \includegraphics[width=0.7\linewidth]{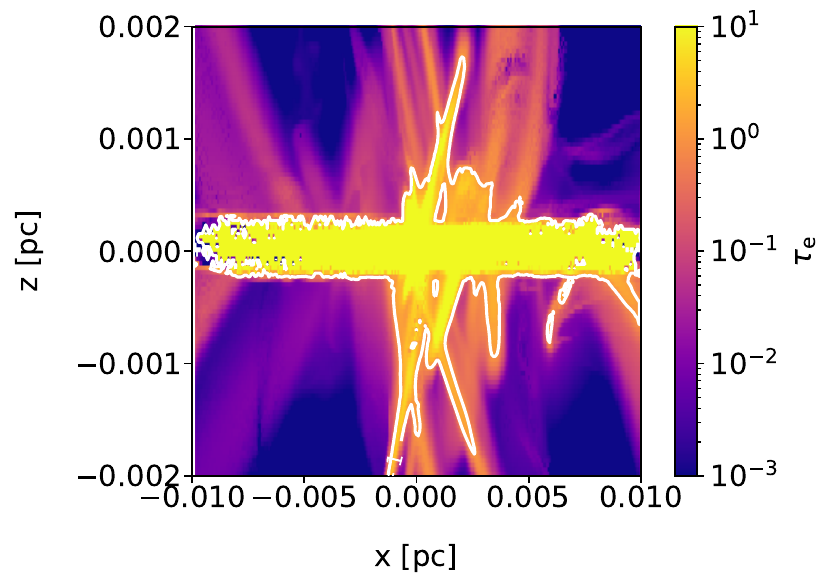} % original: aag0134_panel9.pdf
 \caption{Electron--scattering optical depth map, $T>8000$ K. The white contour represents the photosphere ($\tau_e = 1$). }
 \label{fig:tau_es_map}
 
 % read_binary_phasediagram31.py  panel9
 % plot_tau_temperature_hist2d.py
 %
\end{figure}

Figure \ref{fig:tau_es_map} shows the distribution of the electron-scattering optical depth, $\tau_e$. Here we calculate $\tau_e$ by integrating vertically from the disk midplane ($z=0$). 
{ Note that $\tau_e$ in Figure \ref{fig:tau_es_map} is directly obtained using the result of the hydrodynamic model, while $\tau_e$ in
Figure \ref{fig:sigma-tau} is a parameter to represent the effect of scattering for each emission line (Equation \ref{eq: exp_tail}).}
As indicated by the photosphere (white line), the disk itself is optically thick to electron scattering, but there are ambient layers with $\tau_e \lesssim 0.1$ above the disk.  We suppose that this gas can scatter photons originating from the BLR region.

\subsection{Br$\gamma$ Line Profile}
\begin{figure}[t]
 \centering
 \includegraphics[width=0.55\linewidth]{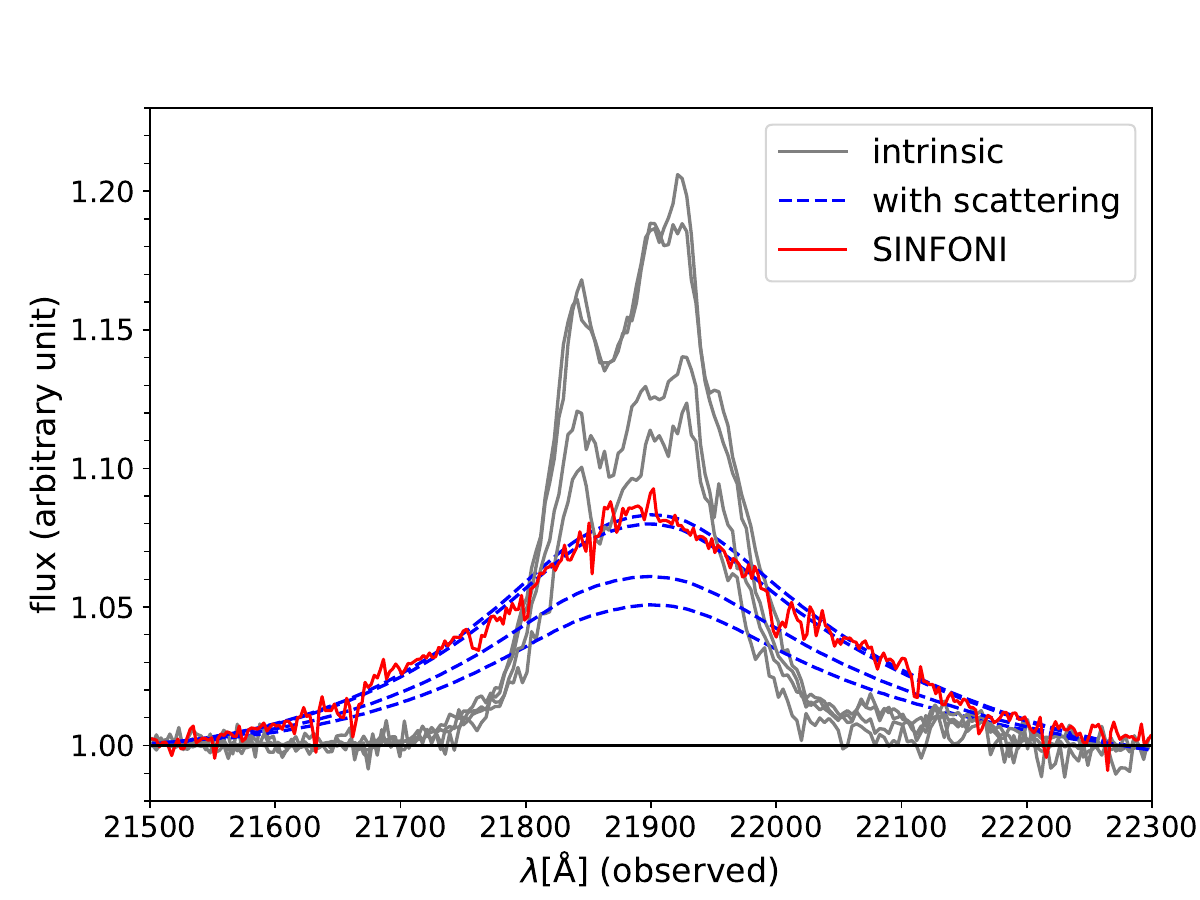} % original: Brg_4files.pdf 
%  plot_Brg_recursive_fitting_EW.py --palett grayred-flat

 \caption{Br$\gamma$ line profile: intrinsic vs.\ scattered model vs.\ SINFONI (red solid line). Gray solid lines are intrinsic line profiles calculated from 
 four different sets of the randomly selected 2000 grid cells. The blue dashed lines are corresponding four profiles with the electron scattering, 
 assuming $\sigma = 2080$ km s$^{-1}$ (Equation (\ref{eq: sigma})) and $\theta_v = 15^\circ$.}
% assuming $\tau_e = 0.1$, $T_e = 5\times 10^5 $K and $\theta_v = 15^\circ$. }
 \label{fig:brg_profile_five}

\end{figure}

{
Figure \ref{fig:brg_profile_five} shows the model \brg profiles (gray solid lines) compared with the SINFONI spectrum (red solid line).
Here we plot four profiles generated from four independent sets of the selected ``BLR candidate cells" (each includes 2000 cells).
 We assume a viewing angle of $\theta_v = 15^\circ$ (the dependence on $\theta_v$ is discussed in Section4). 
 The blue dashed lines are obtained by including electron scattering (Section~3.2), corresponding to the four intrinsic lines.
 Here we assume the dispersion of the electron scattering (Equation \ref{eq: sigma}), $\sigma = 2080$ km s$^{-1}$ 
 for all the lines, and we find that the best match to the observed spectrum are the two lines that are especially the brightest. The other two lines are about 30\%--40\% fainter than the observed spectrum.}
Yet, it is clear that all the intrinsic model lines are evidently narrower than the observed one: the intrinsic profile has FWHM $\approx 200$ \AA ($\approx 2740$ km s$^{-1}$), whereas both the scattered profile and the SINFONI spectrum show FWHM $\approx 400$ \AA. 
The model lines also exhibit substructures that are not seen in the SINFONI spectrum. The two-peak profile commonly seen in all four intrinsic spectra
is caused by the fact that the intrinsic BLR region is a thin rotating disk.

\begin{figure}[H]
 \centering
 \includegraphics[width=0.55\linewidth]{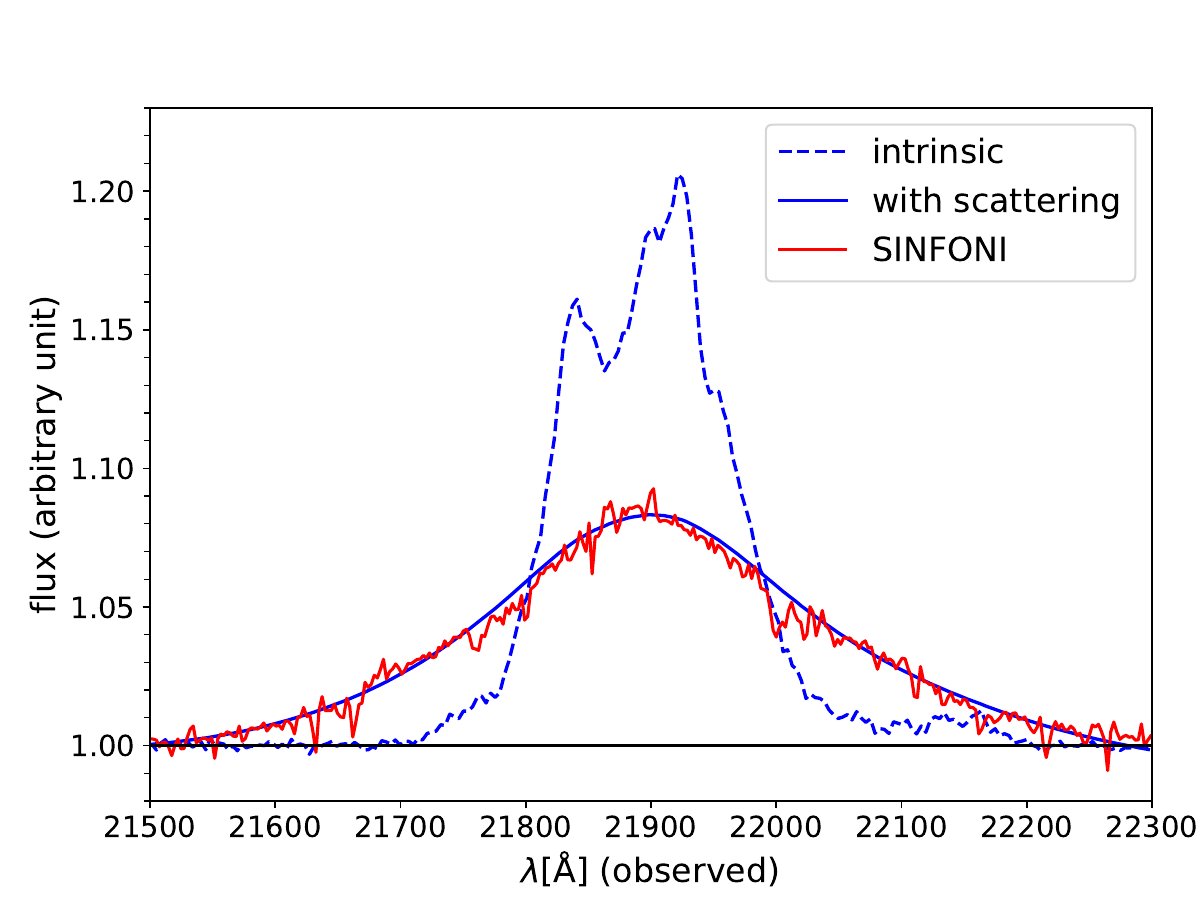} % original: Fig7_v2.pdf

 \caption{Same as Figure \ref{fig:brg_profile_five}, but for one of the data sets, with which 
 the scattered profile well fits the observation. The blue dashed line is the intrinsic \brg profile and the blue 
 solid line is the profile with the electron scattering.}
 \label{fig:brg_profile}
\end{figure}

\begin{figure}[H]
 \centering
 \includegraphics[width=0.55\linewidth]{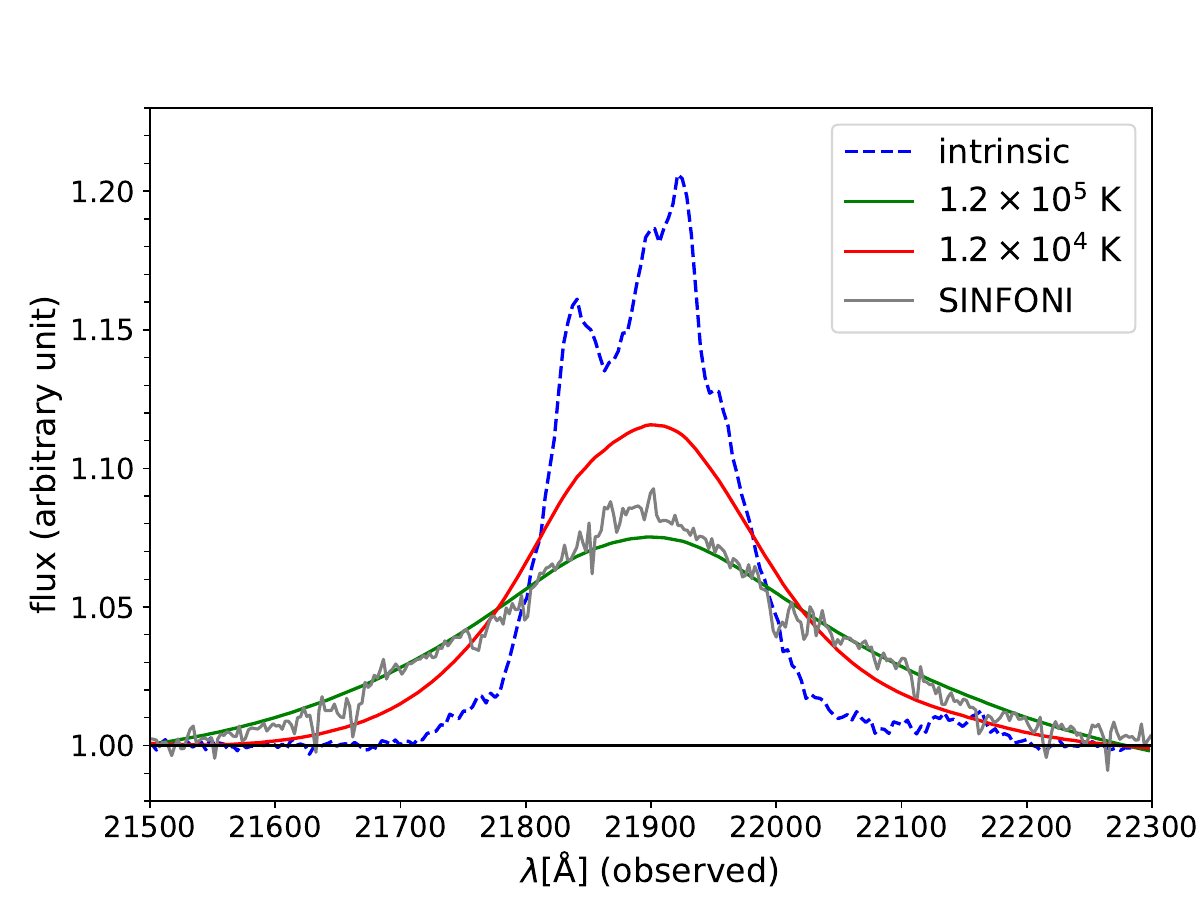} % original: Fig8_v2.pdf
 \caption{Effect of the electron temperature: $T_e = 1.2 \times 10^4$ (red) and $T_e = 1.2 \times 10^5$ K (green), assuming $\tau_e = 0.8$. 
 These values correspond to $\sigma \simeq 930$ km s$^{-1}$ and $2940$ km s$^{-1}$ from Equation (\ref{eq: sigma}).}
 \label{fig:brg_Te}
\end{figure}

{
Figure \ref{fig:brg_profile} shows one of the model profiles that fits the observed Br$\gamma$ profile well.
 One should note that given the simple treatment of the radiative transfer for the scattering (Section~3.2) and the limitation of
 the random sampling for the BLR candidate cells, the best-fit model shown in Figure \ref{fig:brg_profile} does not necessarily represent a unique solution of \brg in NGC~3783.
The equivalent width of the line is sensitive to the subtraction of the continuum, and the continuum levels fluctuate among the cells.
Moreover, considering the distributions of $\tau_e$ and gas temperature $T$ (Fig. \ref{fig:tau_es_map}), the parameters in the best-fit model rather reflect \textit{typical} conditions for the BLR gas and the scattering medium (see Section~\ref{sec: condition}).
Figure~\ref{fig:brg_Te} illustrates the dependence of the line shapes on the electron temperature of the scattering medium (see also Section~4).

}

\begin{figure}[H]
 \centering
 \includegraphics[width=0.5\linewidth]{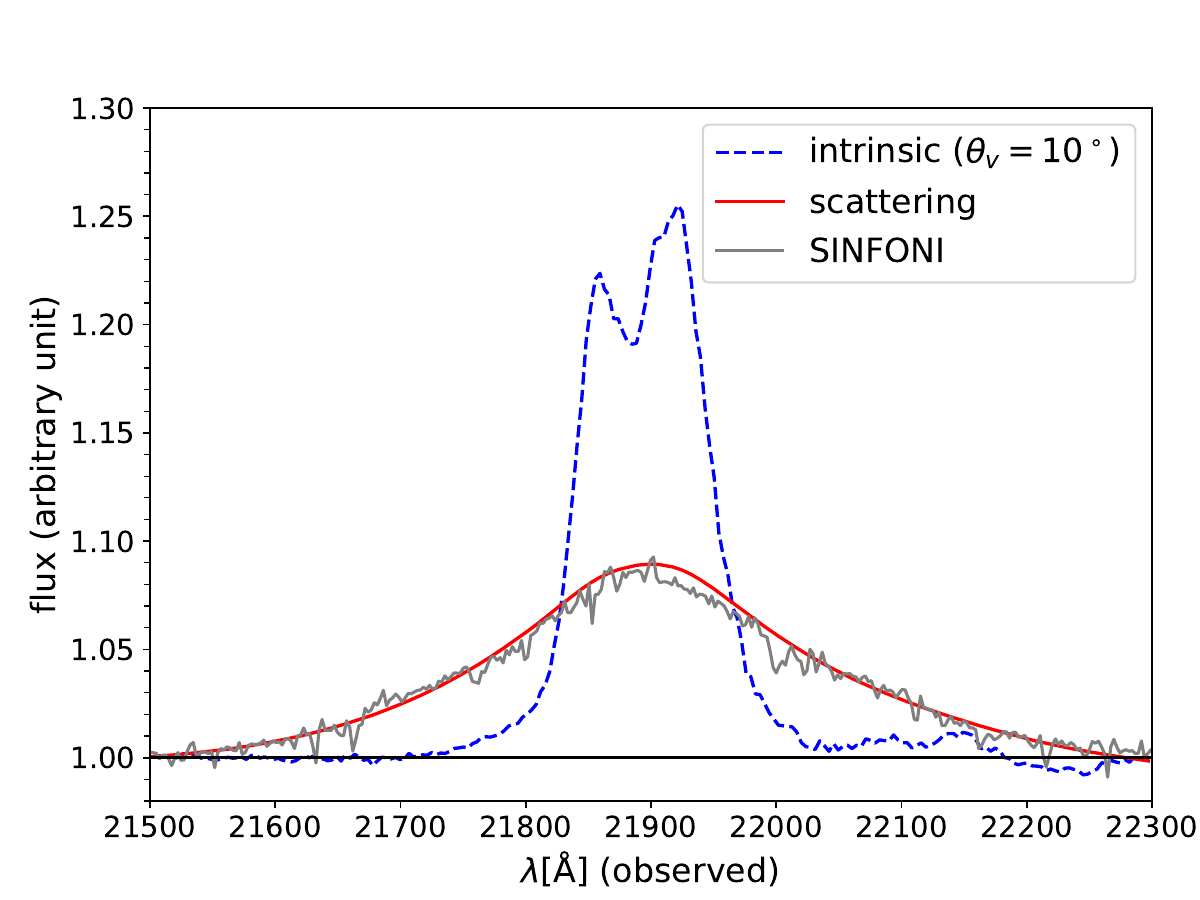} % original: Fig_i10_T05.pdf
 \includegraphics[width=0.5\linewidth]{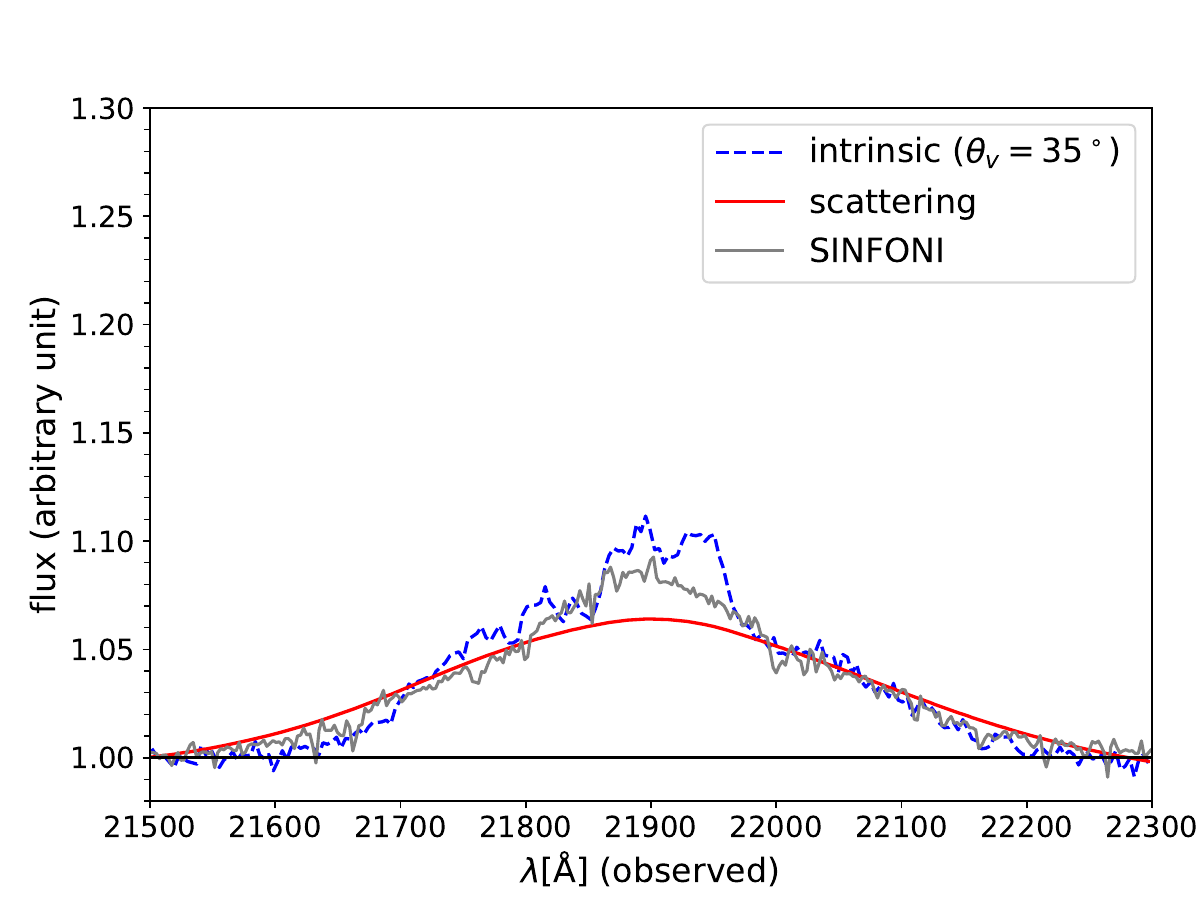} % original: Fig_i35_T05.pdf
 \caption{Same as Figures \ref{fig:brg_profile} and \ref{fig:brg_Te}, but for dependence on the viewing angle: {$\theta_v = 10^\circ$} (top) and 35$^\circ$ (bottom). $\tau_e$ and $T_e$ are the same as in Figure \ref{fig:brg_profile}.  } 
 \label{fig:brg_viewing}
\end{figure}

Figure \ref{fig:brg_viewing} compares Br$\gamma$ line profiles at viewing angles $\theta_v=10^\circ$ and $35^\circ$.
As expected, the intrinsic line is narrower for a nearly face-on view.
The scattered line profiles (red, solid) also depend on viewing angle; however, their differences are less pronounced than for the intrinsic spectra.
Both spectra are smooth and featureless, and the $\theta_v=10^\circ$ model also fits the observations well.
These results suggest that the BLR viewing angle in NGC~3783 is $\sim$10$^\circ$--15$^\circ$, rather than 35$^\circ$.
 GRAVITY observations of NGC~3783 suggested that the BLR clouds are 
distributed in a thick torus-like geometry, and the inclination angle of 
the torus is estimated at $\sim 23^\circ$ \citep[][]{GRAVITY2021a}. 
Given the large ambiguity ($\sim 10^\circ$) in the observationally estimated angle and the insensitive behavior of the line profile with respect to the viewing angle, the GRAVITY observation
is consistent with our model.
One should note, however, that the BLR region does not necessarily have to be a thick `torus' as found in \citet[][]{GRAVITY2021a}. 
Scattering may partially erase not only the velocity structures in the spectra but also the spatial information of the source.
We see the scattered emission mostly from the photosphere (Figure  \ref{fig:tau_es_map}).
In other words, any kind of structure, such as clumps, thin disks, or outflows, may be `blurred' due to the multiple
scattering of the `haze'.
%{However, given the large fluctuations due to sampling (Figure \ref{fig:brg_profile_five}), further investigation is required.}

%%%%%%%%%%%%%%%%%%%%%%%%%%%%%%
%
\section{Discussion}
%
%%%%%%%%%%%%%%%%%%%%%%%%%%%%%%
\subsection{Physical Conditions of the Scattering Medium}
\label{sec: condition}
\begin{figure}[H]
 \centering
 \includegraphics[width=0.6\linewidth]{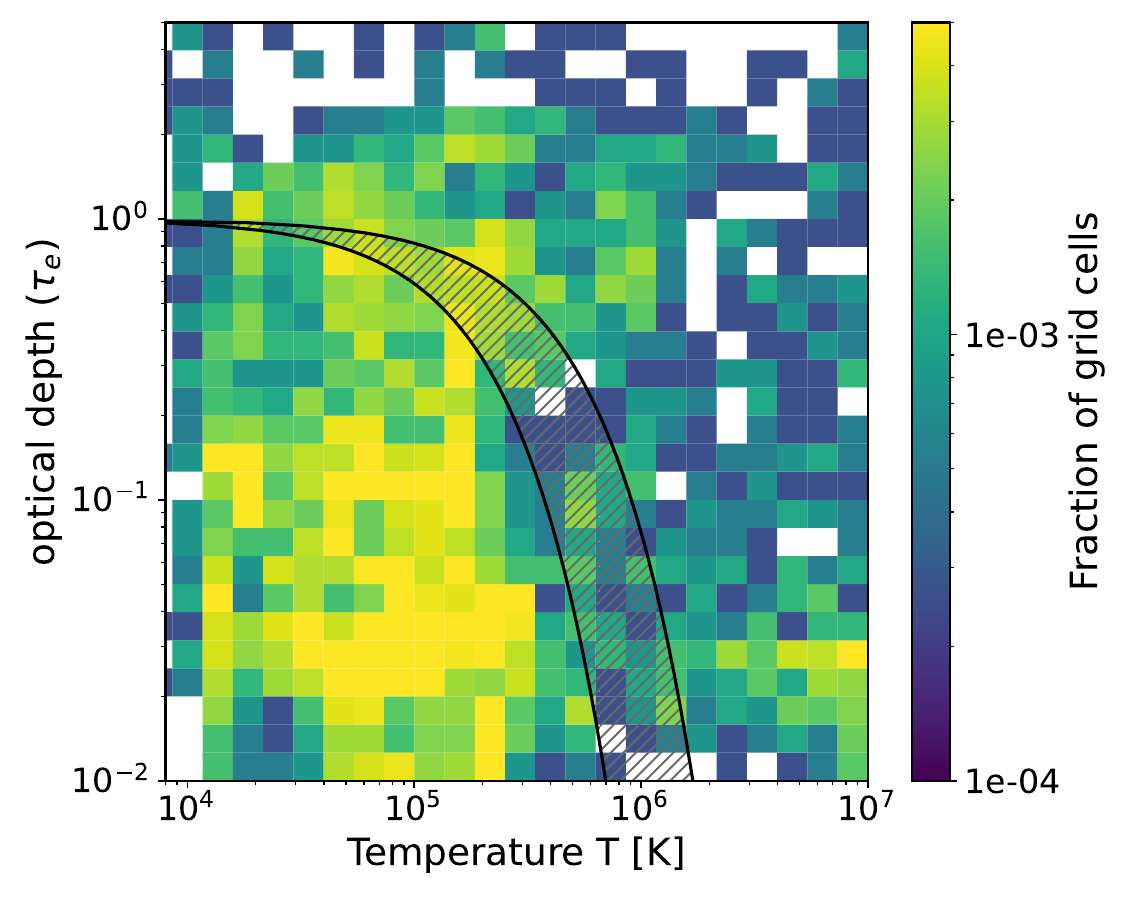} % original: aag0134_tau_vs_T_hist2d_v2.pdf
 \caption{ Volume fraction on a $T$ and $\tau_e$ plane. Upper and lower black thick lines correspond to 
 $\sigma = 2800$ km s$^{-1}$ and $\sigma = 1800$ km s$^{-1}$, respectively.  }
 \label{fig:tau-T}
 
 % read_binary_phasediagram31.py  panel9
 % plot_tau_temperature_hist2d.py
 %
\end{figure}
{
Figures \ref{fig:brg_profile} indicates that the SINFONI \brg profile can be well fit with a convolved 
profile affected by an exponential tail with $\sigma \simeq 2080$ km s$^{-1}$.
In our treatment of electron scattering, as Equations \ref{eq: exp_tail} and \ref{eq: sigma} show, the optical depth $\tau_e$ and the electron temperature $T_e$ are degenerate with respect to $\sigma$.
In Figure \ref{fig:tau-T}, we plot the volume fraction of the hydrodynamic model in the $\tau_e$--$T$ plane.
The hatched region corresponds to  1800 km s$^{-1}$ $< \sigma < 2800$ km s$^{-1}$, for which the SINFONI \brg profile is roughly reproduced
(note that the two models in Figure \ref{fig:brg_Te} correspond to $\sigma =$ 930 and 2940 km s$^{-1}$, and the SINFONI spectrum is closer to the latter).
However, hot gas with small $\tau_e$ (e.g., $\tau_e \lesssim 0.1$ and $T \gtrsim 10^6$ K) should not significantly modify the line profile via electron scattering.
This suggests that a scattering medium with $\tau_e \sim 1$ and $T \lesssim 10^5$ K likely dominates the broadening and smoothing of the intrinsic BLR line profile.
This temperature range corresponds to number densities of $10^4 \lesssim n \lesssim 10^8$ cm$^{-3}$ (see Fig. \ref{fig:phase_dnT}).
These more diffuse, warmer gases (`haze') are not the BLR “clouds” themselves but likely exist around or between them; as a result, the intrinsic spectra emerging from the clouds can be modified by scattering.
}

{ 
While we cannot rule out the possibility that thermal instability 
could form condensed clumps/filaments in diffuse, hot gas, our current 
simulations do not have sufficient resolution to capture such 
fine structures. With our numerical resolution of $3.9 \times 10^{-5}$ pc, we can resolve thermal
instabilities in the "haze" only for $n \lesssim 10^6$ cm$^{-3}$ and $T \gtrsim 10^6$ K.}

\subsection{Scattering, Polarization, and Inflow/outflow}

{
It is well known that scattering of emission from the BLR is a key element of the unified model of AGNs \citep{Antonucci&Miller85}.
The polarization of BLR emission is produced by scattering off electrons in relatively low-temperature ionized gas in the nucleus \citep{Miller1991}.
\citet{Lira2020} studied the polarization-angle profile in the BLR region using the radiative-transfer code STOKES and argued that an “M-type” polarization profile implies that (1) the BLR and the scattering medium are co-spatial, (2) both regions undergo Keplerian rotation, and (3) the scatterer is optically thin in the polar direction but provides sufficient optical depth for photons escaping at low angles with respect to the disk. All of these conditions are consistent with our model.
\citet{Lira2021} subsequently applied their model to NGC~3783 and Mrk 509 and found that the polarization profile shows an “M-shape” in the Balmer lines.
Their best-fit model for NGC~3783 suggested the following parameters: an innermost BLR radius of 0.003 pc; BLR and scatterer scale heights of 0.001 pc; 
a scatterer electron density of $3\times10^{7}\ \mathrm{cm^{-3}}$; and a vertical optical depth of $\tau \simeq 0.07$.
The size and density of the scatterer are close to those suggested in our numerical model. The small optical depth would be reasonable for 
the polarization component.
However, they also argued that the M-type polarization profile in NGC~3783 implies an outflowing scattering medium with velocities of $\sim 4000$--$8000\ \mathrm{km\ s^{-1}}$.
By contrast, we do not find a strong outflow within the central 0.01 pc in our hydrodynamic model, which is reasonable because we assume an Eddington ratio of 10\%
 and fully ionized, dust-free gas. One possible resolution of this discrepancy is that dust survives in a dense, thin disk and radiation pressure on the dusty gas drives a strong outflow, as shown by radiation-hydrodynamic simulations \citep{kudoh2023,kudoh2024}.
}

{
The kinematics of the scatterer remains controversial. For example, \citet{Gaskell2013} claimed that electron and Rayleigh scattering in the BLR and torus can naturally explain the blue-shifted profiles of high-ionization lines. They argued that the blueshift indicates inflow rather than outflow and that this is consistent with velocity-resolved reverberation mapping (e.g., Arp 151; \citealt{bentz2009}). Gaskell’s picture of BLR kinematics—inflow + rotation—is broadly consistent with our model.
As shown in the previous section, the \brg\ profile of NGC~3783 is symmetric, and our model without a steady outflow reproduces this feature well.
}{
Recent VLTI observations have found a strongly outflowing BLR in quasars at $z\approx4$ \citep{GRAVITY+-Collaboration2025-ar}.
If electron scattering of BLR photons also operates in such outflowing AGNs, the observed line profiles may be shaped not only by the intrinsic flow but also by the bulk motion of the scattering material. These are important topics for future self-consistent radiative-transfer simulations that include scattering.
}

{ 
One final comment on the size of the computational box:
due to computational resource limitations,
we focus on the central sub-pc region in this paper. 
However, as we have shown in several papers, radiation from the central source
can drive fountain flows in the outer parsec to 10 pc region \citep[e.g.,][]{Wada2012}.
Inflow from the outer region could affect BLR structures inside the
dust sublimation radius.
However, if we assume a relatively stable AGN (without strong 
perturbations due to e.g., mergers or bars), mass supply to the center is 
likely to occur through the dense gas near the equatorial 
plane. This is indeed what occurs in the axisymmetric calculations of
\citet{kudoh2024}, which cover the range $10^{-4} - 2$ pc, and this is also the case
for the radiation-driven fountain model on the 10 pc scale.
Therefore, the thin disk structure seen in our current calculations may 
not change significantly, even if we account for large-scale fountains
in Seyfert-type AGNs without strong external perturbations.}

\subsection{Implications for Estimating the BH Mass}

{ 
In this paper, we found that the intrinsic emission lines can be narrower than
the observed ones. If this is the case, the mass of SMBH derived from 
the FWHM, assuming gravitational motion of the gas, might be overestimated.}
{ 
The relationship between the observed line profile and the intrinsic 
(i.e., gravity-only) line profile involves degeneracies among BH mass, 
BLR size, degree of electron scattering, Eddington ratio, and viewing 
angle. Within the scope of our theoretical model, we 
cannot make quantitative statements about the contribution of each 
factor.
Qualitatively, for the same BH mass, the observed line profile would 
imply a smaller viewing angle (i.e., closer to face-on). If we fix 
the viewing angle and keep other parameters constant, the intrinsic 
FWHM could be about half the observed value, and the BH mass derived 
assuming only Doppler broadening could be overestimated by a factor of approximately 4.}

{ 
If the intrinsic line width and the scattered line width are 
proportional, or if they do not depend on radius, then the slope of the 
size-velocity width relation, FWHM $\propto L^{-1/4}$, would remain unchanged. 
Conversely, if the slope deviates from $-1/4$ and the intrinsic line 
width is determined primarily by gravity, this could provide indirect evidence that 
the effect of electron scattering is radius-dependent.}

\section{Conclusion}

The broad line spectra in observations are often characterized by a smooth shape. 
This feature was interpreted as the BLR consisting of many discrete cloudlets of ionized gas. 
The large ($>1000$ km s$^{-1}$) velocity width seen in UV--NIR lines in many type 1 AGNs 
implies that the motion of the ionized gas is affected by the gravity of the central supermassive black holes.
The radiative feedback from the accretion disk can also drive the fast motion of the BLR gas.
By fitting the line shape in detail, we could obtain information on the gravitational potential and 
distribution of the BLR clouds. 
Since the emitting regions of the BLR are too small to be resolved even in nearby AGNs, 
the model fitting of lines had to rely on simple toy models, in which the distribution and kinematics of
the BLR cloudlets are assumed. The comparison between observed spectra and models becomes
more reliable if spatial structures of velocity of the lines are obtained.

In this paper we found that BLR spectra—especially near-IR hydrogen lines such as \brg —can be appreciably modified by electron scattering in an electron ``haze" of ionized gas surrounding the intrinsic BLR emission zone, with characteristic electron temperatures and densities of $10^4 \lesssim T_e \lesssim 10^5$ K and $n_e \lesssim 10^{8}$ cm$^{-3}$. This implies that what we observe—both the detailed profile shape and the apparent emissivity distribution—may not be a direct view of the BLR “source” itself. Even if the intrinsic emitter is a geometrically thin, rotating disk (as our fluid calculations naturally predict), its detailed kinematic imprint on the line profile can be washed out by multiple electron scatterings; likewise, any intrinsic substructure or clumpiness can be smeared by propagation through a thin, hot scattering layer. In other words, we may be viewing the BLR through a mild electron mist, so that a large observed line width need not unambiguously imply large bulk velocities (pure rotation, turbulence, or strong outflows).

The true structure of the broad-line region (BLR) has long been debated. Two broad classes of models have been proposed. In one, the BLR consists of a very large number of discrete, nearly noninteracting ionized clouds \citep[e.g.,][]{Krolik1981TwoPhase,Rees1989SmallDense,Baldwin1995LOC}. In the other, the emission arises from a smooth hydrodynamic flow—such as a rotating disk and/or a disk wind—rather than from individually resolved “clouds” \citep[e.g.,][]{Murray1997DiskWind,Proga2000LineDrivenWinds,Pancoast2014DynamicalII,GRAVITY2018Nature3C273}.
The broad similarity of Type 1 AGN line profiles—suggesting limited sensitivity to viewing angle or to large variations in BLR substructure—points to a configuration that, on average, is thicker than a razor-thin disk and may be closer to a thick, rotation-dominated structure (as directly indicated in some sources by interferometry) or even quasi-spherical in its emission weighting \citep{Kollatschny2013ProfileShapes,Pancoast2014DynamicalII,GRAVITY2018Nature3C273}. Constructing a single, self-consistent model—whether cloud-based or flow-based—that simultaneously accounts for these constraints remains challenging.

In this study we demonstrated that, even if the BLR itself consists of the traditionally assumed high-density ionized gas ($n\sim 10^8 - 10^{11}$ cm$^{-3}$), the emergent emission can undergo substantial changes in line shape due to scattering by the surrounding, lower-density and higher-temperature ionized medium. 

Our treatment of the propagation of the emission lines in this paper is, however, phenomenological; ultimately, one must solve the full three-dimensional line radiative transfer including scattering. Although this poses numerous challenges—such as the development of fast radiative-transfer algorithms—we anticipate that, when coupled with three-dimensional hydrodynamic models, such calculations will open a new avenue for probing the origin of the BLR.

\begin{acknowledgments}
We thank P. van Hoof and the \cloudy team for their support and various useful suggestions about \cloudy. 
We also thank the anonymous referee for constructive comments.
We are grateful to P. Papadopoulos, M. Kishimoto, and B. Vander Meulen for their valuable comments on the manuscript.

Numerical computations of the RHD model were performed on a XD50 at the Center for Computational Astrophysics at the National Astronomical Observatory of Japan and Genkai at Research Institute for
Information Technology, Kyushu University. This work was supported by JSPS KAKENHI grant Nos. 25H00671 (K.W.), 23K25911 (T.N.), and the China Manned Space Program (CMS-CSST-2025-A09) (J.S.) and The Fundamental Research Funds for the Central Universities, Peking University (7100604896) (J.S.).
\end{acknowledgments}

%\bibliography{sample631}{}
\bibliography{agn_papers_v3}
%\bibliography{BLR_papers, Paperpile-AGN-Sep02,agn_papers,agn_papers_takasao, reference_baba}
%\bibliography{Paperpile-AGN-Sep02,agn_papers,agn_papers_takasao}
\bibliographystyle{aasjournal}
\end{document}